\documentclass[a4paper]{article}
\usepackage{ISCSLP2026}
\usepackage{multirow}
\usepackage{url}
\usepackage{amsmath}
\usepackage{booktabs}
\usepackage{tikz}
\usetikzlibrary{arrows.meta,positioning,fit,backgrounds}

\title{Modeling, Scaling, and Decoding: Optimizing Controllable Speech Generation with Nonverbal Vocalizations}
\vspace{-3ex}

\name{ Ziyu Zhang, Yun Chen, Taihui Wang, Hanzhao Li, Qicong Xie, Rilin Chen, Zhixian Zhao, Lei Xie}
\vspace{-2ex}
\address{
    Tencent HY Speech Team
}

\vspace{-2ex}

\email{}

\vspace{-5ex}

\begin{document}

\maketitle

\begin{abstract}
\vspace{-1ex}
Controllable synthesis of nonverbal vocalizations (NVVs) is essential for natural and expressive speech, but remains challenging due to their acoustic diversity and imbalanced distribution in existing corpora.
To address these challenges, we develop an NVV-aware DiTAR system that models continuous speech latents, encodes the 16 target NVV categories as dedicated tokens, and adapts stop prediction to distinguish mid-utterance vocalizations from utterance boundaries.
Training begins with large-scale bilingual pre-training on diverse NVV speech, followed by continued supervised fine-tuning on a corpus enhanced through targeted synthetic augmentation and frequency-aware rebalancing.
At inference time, we select the acoustic prompt, tune the LM-guidance and noise-injection scales, and apply Best-of-$N$ sampling with multi-metric selection to reduce generation failures.
The final system achieves an official weighted bilingual score of $62.786$, ranking first in Mandarin, second in English, and first overall among participating systems in Track~2 of the ISCSLP 2026 NVVSpeech Challenge.
Ablation studies show that targeted augmentation benefits underrepresented NVV categories the most, while robust candidate selection requires balancing NVV correctness, lexical fidelity, and perceptual quality.

\end{abstract}

\noindent\textbf{Index Terms}: nonverbal vocalizations, diffusion autoregressive models, data augmentation, best-of-$N$ sampling.

\section{Introduction}
Spoken communication is not carried by words alone. Speakers laugh, sigh,
breathe audibly, clear their throats, and gasp. These nonverbal vocalizations
(NVVs) convey affect, attitude, turn-taking cues, and even physiological
state~\cite{Liao25-NVSpeech,Ye25-NVS38K,Borisov25-NonverbalTTS,Wang25-CapSpeech}.
Conventional text-to-speech (TTS) systems, however, are optimized primarily
for fluent lexical content, and such events are often filtered out during
data curation. Recent zero-shot TTS systems have achieved substantial improvements in
naturalness and speaker similarity through codec language modeling and
large-scale autoregressive generation
~\cite{Wang23-VALLE,Anastassiou24-SeedTTS,Du24-CosyVoice2,Zhou25-IndexTTS2}.
Continuous generative approaches based on diffusion or flow matching further
improve acoustic fidelity and generation flexibility
~\cite{Le23-Voicebox,Chen24-F5TTS,Shen24-NaturalSpeech2,Ju24-NaturalSpeech3,Jia25-DiTAR}.
Despite this progress, existing systems still cannot reliably place a requested
NVV at a specified position with a convincing acoustic realization
~\cite{Zhou25-VoxCPM,Du25-CosyVoice3}.

The NVVSpeech Challenge at ISCSLP 2026 establishes a benchmark for this
capability. In Track~2, systems synthesize speech from transcripts annotated
with one or more of 16 NVV tags, aiming to realize the specified events
naturally while maintaining intelligibility, naturalness, audio quality, and
expressiveness. The revised official protocol combines a Gemini-based LALM
evaluation with subjective listening in Mandarin and English. The
language-specific weighted scores are fused from these two components, and
the overall bilingual score is the average of the Mandarin and English scores.

The lack of an official training set makes NVV data construction a central
challenge, while the substantial acoustic variation within each category calls
for fine-grained modeling. We therefore develop our system along three
dimensions: NVV-aware modeling, data-centric training, and inference-time
decoding.
For model architecture, we adopt DiTAR~\cite{Jia25-DiTAR}, which uses a local
diffusion transformer to autoregressively predict continuous speech latents.
Its patch-wise design builds on advances in diffusion transformers, continuous
autoregressive modeling, and latent generative models
~\cite{Peebles23-DiT,Li24-MAR,Lipman23-FlowMatching,Song21-DDIM,Rombach22-LDM}.
Unlike approaches based on quantized codec tokens
~\cite{Borsos22-AudioLM,Defossez23-EnCodec,Zhang24-SpeechTokenizer}, DiTAR
better preserves the fine-grained acoustic details of NVVs. We introduce the
16 canonical NVV labels as dedicated special tokens and adapt the stop
predictor to distinguish mid-utterance NVV events from utterance boundaries.
To address NVV data scarcity and imbalance, we first pre-train the model on
bilingual speech containing diverse NVV events. We then perform continued SFT
on a corpus enhanced through targeted synthetic augmentation and
tag-frequency-aware rebalancing, increasing exposure to underrepresented
categories~\cite{Cui19-ClassBalanced}.
At inference time, we jointly select the acoustic prompt and tune the guidance
and noise-injection scales using multiple metrics. We further apply
Best-of-$N$ sampling with multi-metric candidate selection to reduce
generation failures.
Experiments show that targeted augmentation provides the largest gains for
severely underrepresented NVV categories. Robust candidate selection also
requires balancing NVV correctness, lexical fidelity, and perceptual quality
rather than optimizing a single metric.

Our contributions are threefold:
(1) we develop an NVV-aware DiTAR with dedicated NVV tokens and adapted stop prediction for controllable NVV synthesis;
(2) we propose a bilingual training recipe combining NVV pre-training, targeted augmentation, frequency-aware rebalancing, and continued SFT; and
(3) we combine decoding-configuration selection with Best-of-$N$ sampling and multi-metric candidate selection to improve generation robustness.
The resulting system ranks first in Mandarin and second in English, achieving first place overall among participating systems in Track~2 of the ISCSLP 2026 NVVSpeech Challenge.



\vspace{-1ex}
\section{System description}
\label{sec:system}

\begin{figure*}[t]
    \centering
    \includegraphics[width=0.8\textwidth]{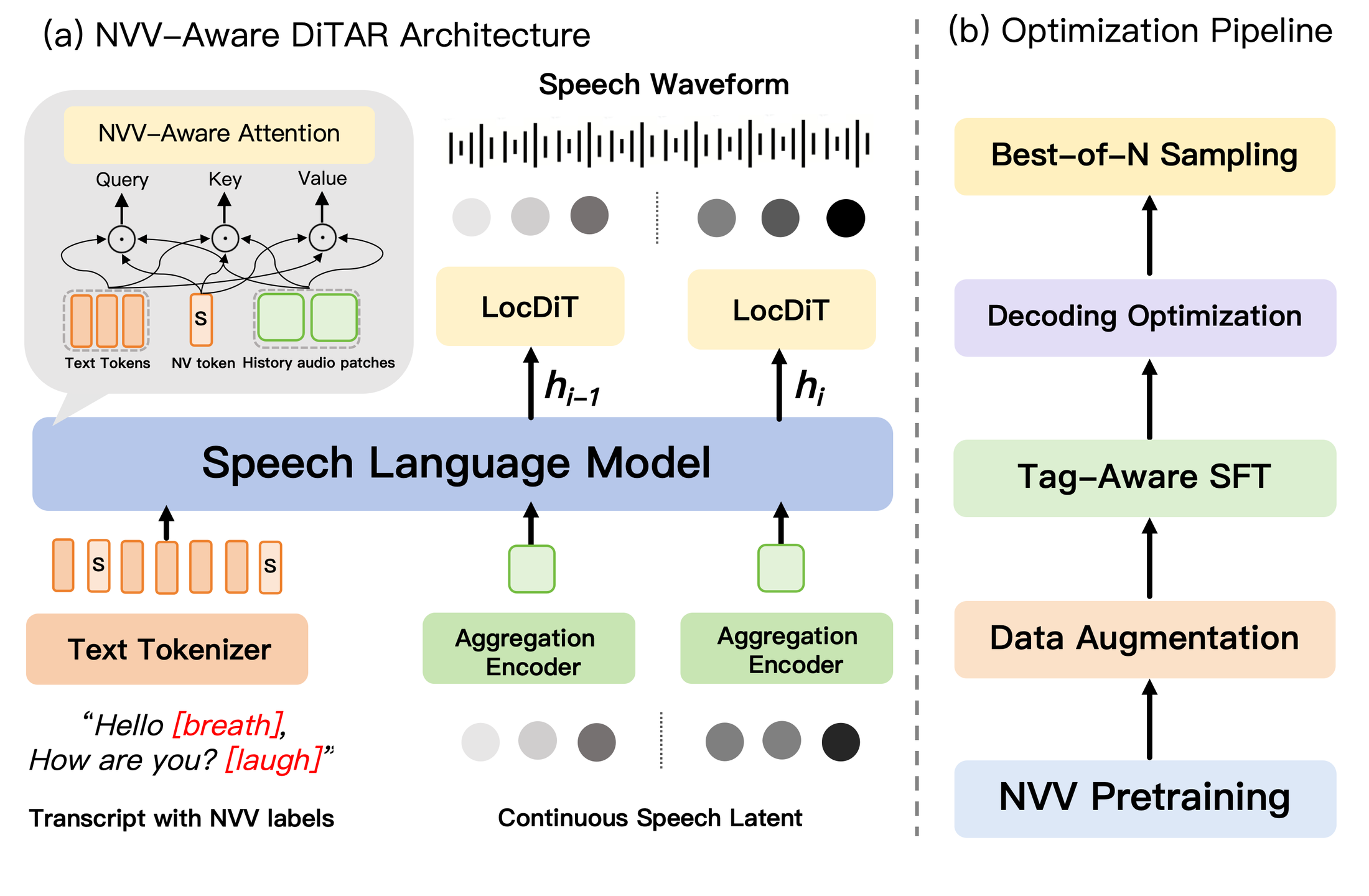}
    \caption{Overview of our NVV-aware speech generation system.
(a) NVV-aware DiTAR, where dedicated NVV tokens condition patch-wise continuous speech generation through causal self-attention.
(b) The training and inference pipeline, including bilingual pre-training, long-tail augmentation, continued SFT, decoding-configuration search, and Best-of-$N$ sampling.}
    \label{fig:system_overview}
\end{figure*}



\subsection{System overview}
Figure~\ref{fig:system_overview} summarizes our system. We first pre-train a
DiTAR backbone on a large-scale bilingual corpus containing both lexical speech
and NVV-annotated speech. We then construct a long-tail-aware trainset
by synthesizing under-represented NVV events, balancing the tag and language
distributions, and continuing supervised fine-tuning (SFT) from the pre-trained
checkpoint. At inference time, we search the best decoding configuration on
the development set and apply best-of-$N$ sampling. The following subsections describe each component.

\subsection{NVV-aware DiTAR Architecture}
\label{sec:backbone}
\textbf{DiTAR backbone.}
As shown in Figure~\ref{fig:system_overview}(a), our system builds on an
in-house $1.5$B-parameter DiTAR model~\cite{Jia25-DiTAR}. DiTAR groups continuous
VAE latents~\cite{Kingma14-VAE} into acoustic patches, summarizes each patch into one embedding, and uses a causal language model to provide patch-level conditions for LocDiT
to generate the next continuous speech patch. This continuous representation
preserves the fine-grained acoustic variations essential for expressive NVV
synthesis. DiTAR also provides the LM-guidance scale $w$, noise-injection scale
$\tau$, and acoustic prompt as native inference controls. We retain these
mechanisms and systematically optimize their configuration on the development
subset, as described in Section~\ref{sec:decoding}.

\textbf{NVV special-token conditioning.}
Our primary architectural adaptation is to register all 16 NVV labels as
dedicated special tokens in the text vocabulary. Each label, such as
\texttt{[laugh]}, is mapped to exactly one token id and a learnable embedding,
rather than being divided into multiple subwords. The LM input is
\begin{equation}
\mathbf{X}=
[\mathbf{e}_1,\ldots,\mathbf{e}_M,
 \mathbf{e}_{\mathrm{audio}},
 \mathbf{a}_0,\ldots,\mathbf{a}_{K-1}],
\label{eq:lm_input}
\end{equation}
where $\mathbf{e}_{1:M}$ contains both ordinary text tokens and NVV special
tokens, $\mathbf{e}_{\mathrm{audio}}$ denotes the start of speech, and
$\mathbf{a}_i$ is an aggregated history-patch embedding. NVV and ordinary text
tokens share the same embedding space and self-attention parameters~\cite{Vaswani17-Transformer,Yang24-Qwen2}. Therefore,
each speech-position query attends to the complete text/NVV prefix, and the
effect of an NVV token is incorporated into $\mathbf{h}_i$ through the standard
attention contribution
$\alpha_{i,\mathrm{NVV}}\mathbf{v}_{\mathrm{NVV}}$. No separate NVV encoder or
explicit event-position input is required.

The retained speech-position states
$\mathbf{H}=[\mathbf{h}_0,\ldots,\mathbf{h}_{K-1}]$ provide the conditions for
next-patch generation. Specifically, $\mathbf{h}_0$ is produced at the
audio-start position, while $\mathbf{h}_i$ for $i>0$ is produced at the
preceding patch position $\mathbf{a}_{i-1}$. Each $\mathbf{h}_i$ is then passed
to LocDiT to generate the next $P$ continuous speech latents.

\textbf{NVV-aware stop prediction.}
Mid-utterance NVVs often introduce pauses or abrupt acoustic changes that may
be confused with the end of an utterance. We therefore adapt the DiTAR stopping
mechanism for NVV generation. We retain its scalar-quantization bottleneck~\cite{Mentzer24-FSQ} and
two-class linear head, but supervise only the final valid patch as
\emph{stop}, while all intermediate patches, including those containing NVVs,
are labelled \emph{continue}. We further use a fixed positive-class weight,
label smoothing, and an un-detached LM hidden state, allowing the stop loss to
shape the semantic representation and distinguish an interrupting NVV from a
true utterance boundary.

\vspace{-2ex}
\subsection{Bilingual pre-training and long-tail-aware SFT}
\vspace{-1ex}

\label{sec:data}

We pre-train the backbone on a large-scale bilingual speech corpus. Its
NVV-containing portion includes public English and Mandarin resources such as
NVSpeech-170K~\cite{Liao25-NVSpeech}, NonVerbalSpeech-38K~\cite{Ye25-NVS38K},
and other NVV-annotated corpora~\cite{Borisov25-NonverbalTTS,Wang25-CapSpeech}.
All NVV annotations are normalized to the 16 challenge categories. This bilingual corpus jointly trains
the newly introduced NVV embeddings and the continuous speech generator while
preserving general TTS capability.

Development-set analysis reveals a highly imbalanced tag distribution, with
limited coverage for events such as \emph{cry}, \emph{snore}, \emph{sneeze},
\emph{yawn}, and \emph{burp}. We therefore use expressive commercial TTS
systems to synthesize English and Mandarin utterances for these weak
categories. NVV tags are placed at sentence-initial, medial, and final
positions, with diverse short contexts and multiple voices to reduce positional
and lexical biases and improve speaker diversity.

We construct the continued-SFT mixture using tag-frequency-aware resampling.
For an utterance whose rarest tag has frequency $f_i$, its repetition factor is
\begin{equation}
r_i=\operatorname{clip}\!\left(
\sqrt{\frac{f_{\max}}{f_i}}\,b_i,\ 1,\ r_{\max}\right),
\label{eq:resample}
\end{equation}
where $b_i$ gives an additional boost to empirically weak tags and $r_{\max}$
prevents excessive duplication. We also upsample English examples to reach the
target English-to-Mandarin ratio. To avoid overrepresenting a few synthetic
speakers or repeated samples, we cap per-speaker contributions and utterance
repetitions and filter excessively long samples. Starting from the bilingual
pre-trained checkpoint, we continue SFT on the rebalanced mixture and select
checkpoints using the local objective metrics described in Section~\ref{sec:eval_setup}, rather than training loss alone.

\vspace{-1ex}
\subsection{Inference-time optimization}
\label{sec:inference}
\label{sec:decoding}
\label{sec:bon}
We optimize inference in two stages: global decoding-configuration search and sample-level Best-of-$N$ selection.

\textbf{Decoding-configuration search.}
The stochastic local diffusion decoder is sensitive to inference settings even
when the model parameters remain unchanged. We therefore evaluate four selected
combinations of the LM-guidance scale $w$ and noise-injection scale $\tau$, as
reported in Table~\ref{tab:decoding_ablation}
~\cite{Ho22-CFG,Sanchez23-CFGLM}. We use the same high-quality reference
acoustic prompt for all configurations. 
We rank these configurations on the development subset using the local
intelligibility, NVV-detection, and audio-quality metrics of
Section~\ref{sec:eval_setup}, and use feedback from 12 official submissions
only to confirm that this ranking transfers to the official metric. Because
the search covers four coarse settings with no per-sample tuning, the risk of overfitting to the evaluation server is limited.

\textbf{Best-of-$N$ selection.}
The selected decoding configuration provides a strong global setting, but
individual generations still vary because of diffusion randomness. For each
tagged transcript, we generate $N=5$ candidates under the same configuration
and rank them automatically using a local multi-metric selector. For each metric, we apply candidate-wise min--max normalization to $[0,1]$,
with a small $\epsilon$ for numerical stability, so that higher values indicate
better candidates. The ASR term is computed from the negative CER or WER. Let $L_i$ denote the
normalized local LALM aggregate using our predefined five-dimension weights,
$D_i$ the normalized score from the fixed NVV detector for realizing the
requested event at the intended position, $R_i$ the normalized ASR-based
lexical-fidelity score, and $Q_i$ the normalized DNSMOS score. We compute
\begin{equation}
C_i=0.40L_i+0.30D_i+0.20R_i+0.10Q_i,
\label{eq:bon_score}
\end{equation}
and select the candidate with the highest $C_i$ as the final waveform. This
sample-level selector uses only local automatic evaluation and does not use
official test feedback to rank individual candidates.

This two-stage strategy first determines a robust global decoding configuration
and then compensates for sample-level generation variance through repeated
sampling. It improves NVV generation reliability without changing the model
parameters, at the cost of additional inference-time computation.


\vspace{-1ex}
\section{Experiments and results}
\label{sec:results}
\subsection{Training and evaluation setup}
\label{sec:eval_setup}

\textbf{Training setup.}
The NVV-aware DiTAR backbone is pre-trained on 100,000 hours of bilingual
speech data. We train the model using the Adam optimizer~\cite{KingmaBa14-Adam}
with a learning rate of $5 \times 10^{-5}$ and a per-GPU batch size of 4 on
32 NVIDIA H20 GPUs, corresponding to a global batch size of 128. The patch size
is set to $P=10$.

\textbf{Evaluation setup.}
We evaluate all systems using local objective metrics and official scores.
Unless otherwise stated, local comparisons use the same evaluation subset,
acoustic prompt, random seed, and decoding configuration. Local evaluation
covers lexical fidelity, NVV realization and localization, and audio quality.
Mandarin CER and English WER are computed using
Paraformer-zh~\cite{Gao22-Paraformer} and
Whisper-large-v3~\cite{Radford23-Whisper}, respectively. A fixed BEATs-based
NVV detector~\cite{Chen23-BEATs} reports precision, recall,
coverage-adjusted F1 (CA-F1), and normalized tag distance (NTD), while
DNSMOS~\cite{Reddy21-DNSMOS} evaluates audio quality. These metrics are used
for model diagnosis, checkpoint selection, and decoding-configuration search.
For sample-level Best-of-$N$ selection, we additionally use a local LALM judge
with the same five-dimensional rubric and our predefined dimension weights;
its aggregate is combined with the objective metrics as defined in
Eq.~\eqref{eq:bon_score}.

The revised official protocol combines a Gemini-based LALM evaluation with
subjective listening. Both components assess NVV accuracy, NVV perceptual
effect, naturalness, audio quality, and expression in Mandarin (ZH) and English
(EN). The official language-specific scores are weighted combinations of the
two evaluation components, and the bilingual score is their arithmetic mean.

\vspace{-1ex}
\subsection{Overall results}
\label{sec:overall}

\begin{table}[!t]
\caption{Objective metrics for progressive system variants. CER and WER
are reported for Mandarin and English, respectively. DNS denotes DNSMOS,
and P, R, and F1 denote precision, recall, and CA-F1.}
\label{tab:main_results}
\centering
\scriptsize
\setlength{\tabcolsep}{1.3pt}
\renewcommand{\arraystretch}{1.05}
\begin{tabular}{@{}lccccccc@{}}
\toprule
\textbf{System}
& \textbf{CER$\downarrow$}
& \textbf{WER$\downarrow$}
& \textbf{DNS$\uparrow$}
& \textbf{P$\uparrow$}
& \textbf{R$\uparrow$}
& \textbf{F1$\uparrow$}
& \textbf{NTD$\downarrow$} \\
\midrule
Official baseline
& 6.267 & 3.159 & 3.026 & 0.769 & 0.251 & 0.569 & 0.073 \\
NVV-aware DiTAR
& 5.603 & 2.949 & 3.126 & 0.816 & 0.263 & 0.570 & 0.071 \\
\quad + Long-tail SFT
& 5.211 & 2.711 & 3.172
& \textbf{0.878} & \textbf{0.275} & 0.574 & 0.069 \\
\quad + Decoding search
& 5.098 & 2.597 & 3.183
& 0.863 & 0.265 & 0.582 & \textbf{0.068} \\
\textbf{\quad + BoN (final)}
& \textbf{5.012} & \textbf{2.194} & \textbf{3.221}
& 0.870 & 0.271 & \textbf{0.589} & 0.069 \\
\bottomrule
\end{tabular}
\end{table}

Table~\ref{tab:main_results} summarizes the progressive development of our
system using local objective metrics. Compared with the official baseline,
the NVV-aware DiTAR reduces Mandarin CER from $6.267\%$ to $5.603\%$ and
English WER from $3.159\%$ to $2.949\%$, while improving DNSMOS from
$3.026$ to $3.126$ and detector precision from $0.769$ to $0.816$.

Long-tail-aware SFT further reduces CER and WER to $5.211\%$ and
$2.711\%$, respectively, while achieving the highest detector precision
and recall of $0.878$ and $0.275$. Decoding-configuration search provides
additional improvements in lexical fidelity and NVV realization, reducing
CER and WER to $5.098\%$ and $2.597\%$ and increasing CA-F1 to $0.582$.
It also achieves the lowest NTD of $0.068$, indicating more accurate NVV
localization.

Best-of-$N$ selection achieves the best CER ($5.012\%$), WER ($2.194\%$),
DNSMOS ($3.221$), and CA-F1 ($0.589$) among the evaluated variants.
Although its NTD slightly increases from $0.068$ to $0.069$, the overall
results show that multi-metric candidate selection provides a favorable
balance among lexical fidelity, audio quality, and NVV realization.

\subsection{Long-tail and per-tag analysis}
\label{sec:per_tag}
We target categories with either limited training coverage or poor source-data
quality. \emph{Cry} and \emph{yawn} are primarily constrained by data scarcity.
This problem is particularly severe for English \emph{cry}, for which few
usable open-source samples are available. In contrast, categories such as
\emph{burp} and \emph{laugh} have more training samples, but many exhibit
limited acoustic diversity or noticeable synthetic artifacts. We therefore
supplement these weak categories with targeted synthetic data and increase
their exposure during continued SFT through frequency-aware resampling.

\begin{table}[t]
  \caption{Targeted augmentation for weak NVV categories. Original/Added denote the numbers of initial/synthesized utterances, and $\Delta S$ denotes the change from NVV-aware DiTAR to long-tail-aware SFT.}

  \label{tab:per_tag}
  \centering
  \scriptsize
  \setlength{\tabcolsep}{2.4pt}
  \renewcommand{\arraystretch}{1.03}
  \begin{tabular}{@{}lccccc@{}}
    \toprule
    \multirow{2}{*}{\textbf{Tag}} &
    \multicolumn{2}{c}{\textbf{ZH utterances}} &
    \multicolumn{2}{c}{\textbf{EN utterances}} &
    \multirow{2}{*}{\textbf{$\Delta S_{\mathrm{ZH/EN}}$}} \\
    \cmidrule(lr){2-3}\cmidrule(lr){4-5}
    & \textbf{Original} & \textbf{Added} &
      \textbf{Original} & \textbf{Added} & \\
    \midrule
    cry     & 5,424  & 500   & 37    & 1,000 & $+4.63/+23.30$ \\
    snore   & 3,728  & 200   & 1,023 & 400   & $+1.50/+0.78$ \\
    sneeze  & 3,657  & 500   & 1,024 & 400   & $-1.15/-1.98$ \\
    yawn    & 252    & 1,000 & 17    & 1,000 & $+3.35/+2.95$ \\
    burp    & 4,328  & 400   & 1,726 & 400   & $+7.30/+2.30$ \\
    laugh   & 29,077 & 500   & 3,329 & 500   & $+3.80/+7.00$ \\
    \bottomrule
  \end{tabular}
\end{table}

As shown in Table~\ref{tab:per_tag}, English \emph{cry} achieves the largest
gain of $23.30$ points, while Mandarin \emph{cry} improves by $4.63$ points.
The English improvement is mainly driven by higher NVV accuracy and perceptual
effect, suggesting that the later system produces NVV events more reliably
and makes them more clearly perceptible. Similarly, \emph{yawn}, \emph{laugh},
and \emph{burp} improve in both Mandarin and English, with the Mandarin
\emph{burp} category exhibiting a particularly substantial gain of $7.30$
points. However, the changes are not uniformly positive. \emph{Snore} remains
nearly unchanged, while \emph{sneeze} degrades in both languages. These results
indicate that increasing sample count is most effective when missing coverage
is the main bottleneck. 
When the synthesized samples lack
naturalness, speaker diversity, or acoustic variation, additional data may
introduce source-specific biases rather than improve generalization.

\vspace{-1ex}
\subsection{Ablation Study}

\textbf{Decoding configuration.}
Table~\ref{tab:decoding_ablation} examines the effects of the LM-guidance scale
$w$ and noise-injection scale $\tau$ using a fixed model checkpoint. Among the
evaluated configurations, $w=2.0$ and $\tau=0.9$ achieves the best lexical
fidelity, reducing CER and WER to $5.098\%$ and $2.597\%$, respectively. It
also obtains the highest CA-F1 of $0.582$, while maintaining a DNSMOS score
close to the best result. This suggests that a higher noise-injection scale
improves both content generation and NVV realization under a moderate guidance
scale.

The different metrics nevertheless reveal a trade-off. The configuration
$w=2.5,\tau=0.7$ achieves the lowest NTD, indicating more accurate event
placement, but performs worse in CER, WER, and DNSMOS. Similarly, the highest
DNSMOS is obtained with $w=2.0,\tau=0.7$, although the difference from
$w=2.0,\tau=0.9$ is small. We therefore select $w=2.0,\tau=0.9$ as the most
balanced configuration rather than optimizing any single metric.
\begin{table}[t]
  \centering
  \caption{Decoding-configuration ablation on the development subset. All rows
  use the same checkpoint.}
  \label{tab:decoding_ablation}
  \scriptsize
  \setlength{\tabcolsep}{2.2pt}
  \renewcommand{\arraystretch}{1.03}
  \begin{tabular}{@{}ccccccc@{}}
      \toprule
      \textbf{$w$} & \textbf{$\tau$} &
      \textbf{CER$_{\mathrm{ZH}}\downarrow$} &
      \textbf{WER$_{\mathrm{EN}}\downarrow$} &
      \textbf{DNSMOS$\uparrow$} &
      \textbf{CA-F1$\uparrow$} &
      \textbf{NTD$\downarrow$} \\
      \midrule
      2.5 & 0.7 & 6.217 & 3.017 & 3.161 & 0.569 & \textbf{0.058} \\
      1.5 & 0.7 & 5.998 & 2.614 & 3.177 & 0.543 & 0.062 \\
      2.0 & 0.7 & 5.741 & 2.661 & \textbf{3.188} & 0.558 & 0.065 \\
      2.0 & 0.9 & \textbf{5.098} & \textbf{2.597} & 3.183 & \textbf{0.582} & 0.068 \\

    \bottomrule
  \end{tabular}
\end{table}

\textbf{BoN candidate selection.}
Table~\ref{tab:bon_selector} compares candidate-selection strategies using the
same pool of five samples. The results show that single-metric selectors exhibit
clear metric-specific biases. DNSMOS-only selection achieves the highest audio
quality, but substantially degrades CER and WER. In contrast, NVV-detector-only
selection obtains the highest CA-F1, but produces the lowest DNSMOS and a larger
position error. Random selection also fails to improve lexical fidelity
consistently, confirming that the benefit of BoN does not arise merely from
generating more candidates.

The multi-metric selector achieves the lowest CER and WER, while improving
CA-F1 and DNSMOS over single-sample generation. Although its CA-F1 is slightly
lower than that of the detector-only selector, it avoids the pronounced loss in
audio quality and lexical fidelity caused by optimizing the detector score
alone. These results demonstrate that jointly considering lexical fidelity,
NVV realization, and audio quality provides a better overall trade-off than
selecting candidates according to any individual metric.

\begin{table}[t]
    \centering
    \caption{BoN selection ablation with five candidates.}
    \label{tab:bon_selector}
    \scriptsize
    \setlength{\tabcolsep}{2.4pt}
    \renewcommand{\arraystretch}{1.03}
    \begin{tabular}{@{}lccccc@{}}
      \toprule
      \textbf{Selection strategy} &
      \textbf{CER$_{\mathrm{ZH}}\downarrow$} &
      \textbf{WER$_{\mathrm{EN}}\downarrow$} &
      \textbf{DNSMOS$\uparrow$} &
      \textbf{CA-F1$\uparrow$} &
      \textbf{NTD$\downarrow$} \\
      \midrule
      Single sample      & 5.743 & 3.148 & 3.157 & 0.542 & \textbf{0.059} \\
      Random selection   & 6.311 & 3.277 & 3.161 & 0.581 & 0.067 \\
      DNSMOS only        & 6.012 & 3.480 & \textbf{3.229} & 0.543 & 0.061 \\
      NVV detector only  & 5.977 & 2.792 & 3.102 & \textbf{0.591} & 0.068 \\
      Multi-metric       & \textbf{5.012} & \textbf{2.194} & 3.221 & 0.589 & 0.069 \\
      
      \bottomrule
    \end{tabular}
\end{table}

\begin{table}[!t]
\caption{Official Track~2 results.}
\label{tab:official_results}
\centering
\scriptsize
\setlength{\tabcolsep}{3.0pt}
\renewcommand{\arraystretch}{1.05}
\begin{tabular}{@{}lccc@{}}
\toprule
\textbf{System}
& \shortstack{\textbf{LALM}\\
              \textbf{ZH / EN / Bi.}}
& \shortstack{\textbf{Subjective}\\
              \textbf{ZH / EN / Bi.}}
& \shortstack{\textbf{Final}\\
              \textbf{ZH / EN / Bi.}} \\
\midrule

Ours
& \shortstack{76.22\\
              69.66\\
              72.94}
& \shortstack{2.838$\pm$0.096\\
              3.156$\pm$0.099\\
              2.997$\pm$0.070}
& \shortstack{61.775\\
              63.797\\
              62.786} \\

Official baseline
& \shortstack{77.62\\70.30\\73.96}
& \shortstack{2.714$\pm$0.089\\
              3.072$\pm$0.091\\
              2.893$\pm$0.066}
& \shortstack{59.700\\
              63.163\\
              61.431} \\

\bottomrule
\end{tabular}
\end{table}

Table~\ref{tab:official_results} reports the revised official results. Our
system obtains bilingual averages of $72.94$ in LALM evaluation and
$2.997\pm0.070$ in subjective listening. Their combination yields final scores
of $61.775$ for Mandarin, $63.797$ for English, and $62.786$ overall, exceeding
the baseline by $2.075$, $0.634$, and $1.355$ points, respectively. It ranks
first in Mandarin, second in English, and first overall among participating
systems, confirming the complementary value of automatic and human evaluation.

\vspace{-1ex}
\subsection{Discussion}
\label{sec:discussion}
Three findings emerge. First, objective, LALM-based, and subjective metrics
provide complementary evidence. Second, targeted augmentation is most effective
for data-scarce categories, whereas low-diversity synthetic data yields
inconsistent gains. Third, decoding optimization and multi-metric BoN improve
performance without model updates, albeit with extra inference cost. The
second-place English rank and lower English LALM score indicate that English
NVV control remains an important direction.

\vspace{-1ex}
\section{Conclusion}
We presented an NVV-aware DiTAR system combining continuous-latent modeling,
long-tail-aware training, decoding search, and multi-metric Best-of-$N$
selection. It achieved an official weighted bilingual score of $62.786$ and
ranked first overall among participating Track~2 systems. Future work will
improve English NVV data and event-level selection metrics.

\bibliographystyle{IEEEtran}
\bibliography{mybib_complete}

\end{document}